\documentclass[pra,twocolumn,superscriptaddress,showpacs,amsmath,amssymb]{revtex4-1}
\usepackage{graphicx}
\usepackage{bm}
\usepackage{amssymb}
\usepackage{amsmath}
\usepackage{dcolumn}
\usepackage{color}

\newcommand{\ket}[1]{| #1 \rangle}
\newcommand{\bra}[1]{\langle #1 |}
\newcommand{\ketbra}[2]{| #1 \rangle \langle #2 |}
 
\def\H{{\rm H}}
\def\V{{\rm V}}

\def\00{\H\V}
\def\11{\V\H}

\begin{document}
\title{
High-fidelity entanglement swapping and generation of three-qubit GHZ state
using asynchronous telecom photon pair sources 
}

\author{Yoshiaki~Tsujimoto}
\affiliation{Graduate School of Engineering Science, Osaka University,
Toyonaka, Osaka 560-8531, Japan}
\author{Motoki~Tanaka}
\affiliation{Graduate School of Engineering Science, Osaka University,
Toyonaka, Osaka 560-8531, Japan}
\author{Nobuo~Iwasaki}
\affiliation{Graduate School of Engineering Science, Osaka University,
Toyonaka, Osaka 560-8531, Japan}
\author{Rikizo~Ikuta}
\affiliation{Graduate School of Engineering Science, Osaka University,
Toyonaka, Osaka 560-8531, Japan}
\author{Shigehito~Miki}
\affiliation{Advanced ICT Research Institute, National Institute of 
Information and Communications Technology (NICT), Kobe 651-2492, Japan}
\author{Taro~Yamashita}
\affiliation{Advanced ICT Research Institute, National Institute of 
Information and Communications Technology (NICT), Kobe 651-2492, Japan}
\author{Hirotaka~Terai}
\affiliation{Advanced ICT Research Institute, National Institute of 
Information and Communications Technology (NICT), Kobe 651-2492, Japan}
\author{Takashi~Yamamoto}
\affiliation{Graduate School of Engineering Science, Osaka University,
Toyonaka, Osaka 560-8531, Japan}
\author{Masato~Koashi}
\affiliation{Photon Science Center, The University of Tokyo, 
Bunkyo-ku, 113-8656, Japan}
\author{Nobuyuki~Imoto}
\affiliation{Graduate School of Engineering Science, Osaka University,
Toyonaka, Osaka 560-8531, Japan}

\begin{abstract}
We experimentally demonstrate a high-fidelity entanglement swapping 
and a generation of the Greenberger-Horne-Zeilinger~(GHZ) state using polarization-entangled photon pairs 
at telecommunication wavelength 
produced by spontaneous parametric down conversion with continuous-wave pump light. 
While spatially separated sources asynchronously emit photon pairs, 
the time-resolved photon detection guarantees the temporal indistinguishability of photons 
without active timing synchronizations of pump lasers and/or adjustment of optical paths. 
In the experiment, photons are sufficiently narrowed by 
fiber-based Bragg gratings with the central wavelengths of 1541~nm $\&$ 1580~nm, and 
detected by superconducting nanowire single-photon detectors with low timing jitters. 
Observed fidelities are $0.84\pm0.04$ and $0.70\pm0.05$ for the entanglement swapping and generation of the GHZ state, 
respectively. 
\end{abstract}

\maketitle

\section{Introduction}
Entanglement swapping~\cite{PhysRevLett.71.4287} is an 
entangling operation on two independent photons and 
a key technique for implementing various quantum information processing such as 
quantum repeaters~\cite{RevModPhys.83.33}, and 
quantum computation~\cite{knill2001scheme}. 
In order to perform such
tasks successfully, indistinguishability of the independently 
generated photons is of importance. 
Spontaneous parametric down conversion~(SPDC) 
is a standard method to generate entangled photon pairs 
and many experiments of entanglement swapping utilizing SPDC have been 
demonstrated~\cite{PhysRevLett.80.3891, PhysRevLett.88.017903, PhysRevA.71.050302, 
PhysRevLett.96.110501, 1367-2630-11-3-033008, jin2015highly}. 
In such experiments, SPDC photons are generated by using ultrafast pulsed lasers, and 
temporal indistinguishability of the photons is provided through 
a timing synchronization of the photon generation and/or the precise stabilization of the optical path lengths. 
However, when we look at a long-distance quantum communication, 
precise timing synchronization of the distant lasers becomes challenging 
since the overall timing instability must be within the coherence time. 
This difficulty is removed by using continuous wave~(cw) 
pumped photon pair 
sources and the coincidence detection with a temporal resolution much shorter 
than the coherence time of the photons, 
which is realized by increasing coherence time of photons 
with narrow bandpass filters and/or employing photodetectors with low timing jitters~\cite{halder2007entangling, 1367-2630-10-2-023027, PhysRevA.80.042321, 6985715, 1608.04909}. 
The advantage of this method 
is the non-necessity of any active 
timing synchronizations of the photon sources. 
In addition, the prerequisite of the narrow band width of photon pairs allows 
the enchantment of the efficiency by the dense frequency multiplexing. 

The entanglement swapping using asynchronous photon pair source is firstly demonstrated 
by Halder $\it{et}~\it{al.}$~\cite{halder2007entangling}, in which 
they utilized energy-time entangled photon pairs at telecom wavelengths generated by cw-pumped SPDC 
and photodetectors with low timing jitters. 
In the demonstration, an observed visibility and a four-fold count rate 
are 0.63 $\pm$ 0.02 and 5~counts/hour, respectively, 
which are much smaller than those observed with a timing synchronization 
of pulse-pumped SPDC~\cite{PhysRevLett.88.017903, PhysRevA.71.050302, 
PhysRevLett.96.110501, 1367-2630-11-3-033008, jin2015highly}. 
Higher efficiency and visibility will be desirable for performing various kinds of applications 
such as quantum repeaters~\cite{RevModPhys.83.33}, quantum relays~\cite{PhysRevA.66.052307, collins2005quantum}, 
measurement-device-independent quantum key distribution~(QKD)~\cite{PhysRevLett.108.130503, pirandola2015high, Scherer:11} 
and distributed quantum computation~\cite{PhysRevA.89.022317}. 

 \begin{figure*}[t]
 \begin{center}
\scalebox{0.47}{\includegraphics{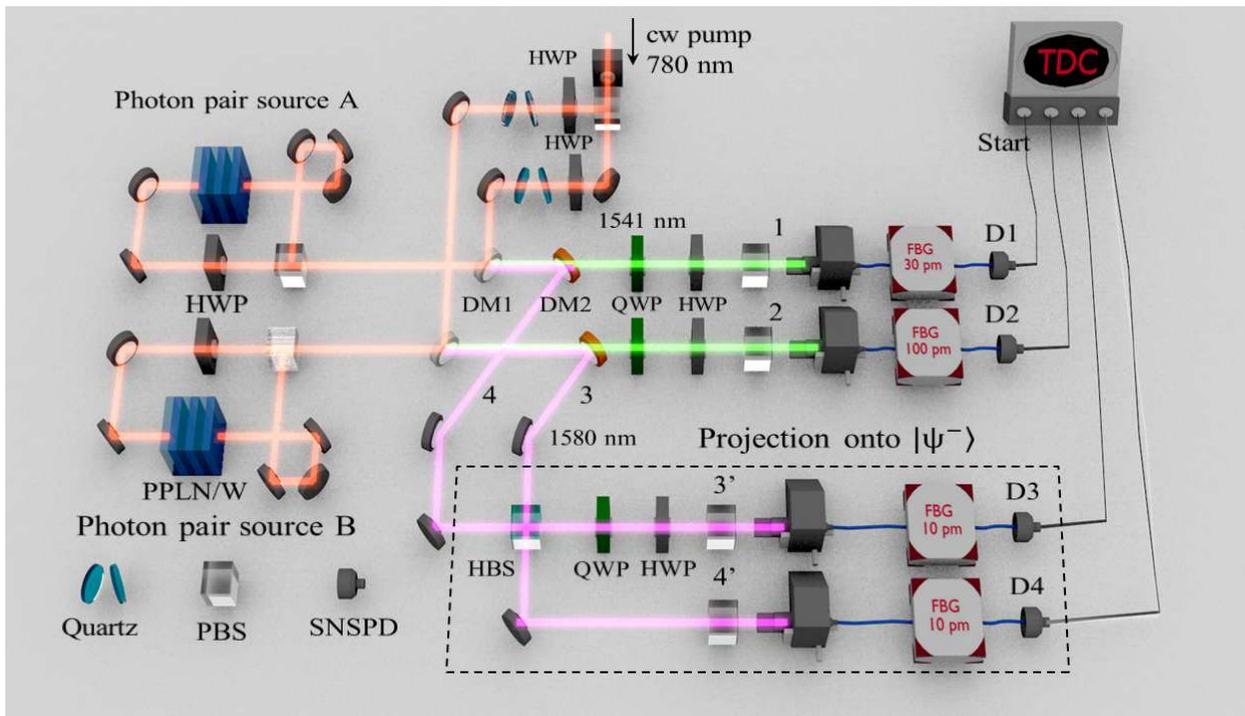}}
  \caption{(Color online)~
  Experimental setup. The cw pump beam at 780~nm is obtained by the 
  second-harmonic generation based on a periodically-poled lithium niobate waveguide~(PPLN/W). An entangled photon pair at 1541~nm and 1580~nm 
  is generated by PPLN/W in a Sagnac configuration. After projecting the polarization state of photons at 1580~nm, 
  the photons at 1541~nm become entangled. 
     \label{fig:swappingsetup}}
 \end{center}
\end{figure*}

In this paper, we show high-fidelity and extended 
entanglement manipulation using telecom-band asynchronous polarization entangled photon pairs.
First we demonstrate high-fidelity entanglement swapping 
using polarization entangled photon pairs 
generated by cw-pumped SPDC with high-resolution photon detectors. 
We performed quantum state tomography~(QST) on swapped photon pairs 
and reconstructed its density operator.
The observed fidelity is $0.84\pm0.04$, which is 
much higher than the previously reported value, and 
as high as those observed by using pulse-pumped SPDC. 
Second, we for the first time demonstrated 
generation of a telecom-band three-qubit Greenberger-Horne-Zeilinger~(GHZ) state 
using asynchronous sources, 
and observed a fidelity of $0.70\pm0.05$, which is 
applicable to not only 
the fundamental test of nonlocality~\cite{RevModPhys.81.865} but also
quantum communication using multipartite entanglement 
such as secret sharing~\cite{PhysRevLett.95.200502} and quantum cryptography~\cite{RevModPhys.74.145}. 
In our experiments, the efficiencies are also highly improved compared to the previous experiment~\cite{halder2007entangling}. 
We obtain the four-fold coincidence rates of 28~counts/hour and 131~counts/hour 
for entanglement swapping and GHZ state generation, respectively. 
Furthermore, by enlarging the width of each coincidence window to be as large as the coherence time of SPDC photons, 
we still kept a high fidelity of 0.75 $\pm$ 0.02 for entanglement swapping with a four-fold coincidence rate of as high as 100~counts/hour. 
These results pave the way for high-quality and efficient 
photonic quantum information processing in cw regime. 

\section{Experiment}
\subsection{Experimental setup}
The experimental setup is shown in Fig.~\ref{fig:swappingsetup}. 
A cw pump beam at 780~nm is obtained by 
second-harmonic generation of light at 1560~nm from an external 
cavity diode laser with a linewidth of 1.8~kHz~\cite{1608.04909}. 
At photon pair source~A, 
an entangled photon pair $\ket{\phi^+}=(\ket{\H\H}+\ket{\V\V})/\sqrt{2}$ 
at 1541~nm and 1580~nm is generated by a 40-mm-long and 
type-0 quasi-phase-matched periodically-poled lithium niobate waveguide~(PPLN/W) 
in a Sagnac configuration with a polarizing beamsplitter~(PBS), 
where $\ket{\H}$ and $\ket{\V}$ represent horizontal~(H) and vertical~(V) polarization states of a photon. 
The pump beam is removed from the generated SPDC photons by a dichroic mirror~1~(DM1). 
DM2 divides spatial modes of the two photons at 1541~nm and 1580~nm into modes~1 and~4, respectively. 
Similarly, we prepared the other entangled photon pair at 1541~nm in mode~2 and 1580~nm in mode~3
at photon pair source~B~in Fig.~\ref{fig:swappingsetup}. 
Photons at 1580~nm in mode~3 and mode~4 are mixed by a half beamsplitter~(HBS).
The coincidence detection between mode~3' with V-polarization and mode 4' with H-polarization 
is regarded as a projection of a photon pair in modes~3 and 4 into the singlet state $\ket{\psi^-}=(\ket{\H\V}-\ket{\V\H})/\sqrt{2}$ ideally. 
As a result, the polarization state in modes~1 and 2 also becomes $\ket{\psi^-}$. 
In order to improve the indistinguishability of the photons in modes~3' and 4', 
the photons are filtered by fiber-based Bragg gratings~(FBGs) with bandwidths of 
10~pm followed by superconducting nanowire single-photon detectors~(SNSPDs) whose timing jitter is $\tau_j=85$~ps each~\cite{Miki:13}. 
The photons in modes~1 and 2 are also filtered by FBGs with bandwidths of 30~pm and 
100~pm, respectively, which plays a role of avoiding saturation of the single counts of SNSPDs. 
The photon's coherence time $\tau_c$ is estimated by 
the temporal distribution of two-fold coincidence events of the photon pairs. 
Approximating it by a Gaussian and assuming that it is the convolution of 
three Gasussians with widths $\tau_j$, $\tau_j$, and $\tau_c$, we estimated $\tau_c$ to be 
about 230 ps FWHM, 
which satisfies the condition $\tau_{c}\gg\tau_{j}$ 
for a high-visibility interference~\cite{halder2007entangling, 1608.04909}. 
The electric signal from D1 is connected to a time-to-digital converter~(TDC) as
a start signal, and the electric signals from D2, D3 and D4 are used as the stop signals of the TDC. 
We collect all of timestamps of the stop signals for every start signal with time slot of 1~ps.  
We postselect the records of the three stop signals within time windows $\tau_\mathrm{w}$. 
If there is at least one detection event in each stop signal, we regard this
event as a four-fold coincidence event. 

\subsection{Entanglement swapping}
\begin{figure}[t]
 \begin{center}
 \includegraphics[width=\columnwidth]{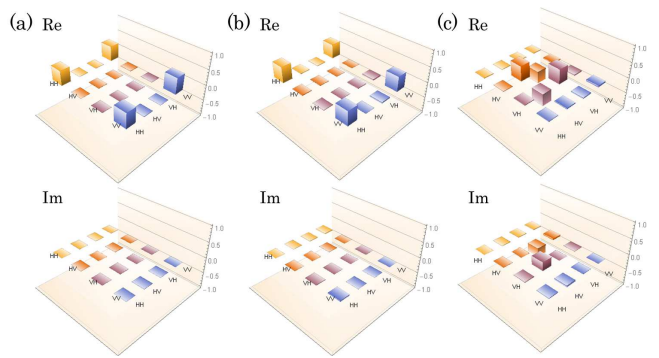}
  \caption{(Color online)~
  The real parts and imaginary parts of the density operators of (a)~$\hat{\rho}_A$, (b)~$\hat{\rho}_B$ and (c)~$\hat{\rho}_{\rm{swap}}$. 
     \label{fig:matrix}}
 \end{center}
\end{figure}

Before performing entanglement swapping, 
we first characterized the initial entangled photon pairs from photon pair source~A~($\hat{\rho}_A$) and photon pair source~B~($\hat{\rho}_B$) 
by measuring the two-fold coincidence count between D1 $\&$ D3, and D2 $\&$ D3, respectively. 
By performing the QST and diluted maximum-likelihood algorithm~\cite{PhysRevA.75.042108}, 
we reconstructed the density operators as shown in Fig.~\ref{fig:matrix}~(a) and~(b). 
Observed fidelities defined by $\bra{\phi^+}\hat{\rho}_A\ket{\phi^+}$ and $\bra{\phi^+}\hat{\rho}_B\ket{\phi^+}$ were 0.963 $\pm$ 0.004 
and 0.931 $\pm$ 0.004, respectively, which implies that the two photon pairs are highly entangled. 
The detection rates of  $\hat{\rho}_A$ and $\hat{\rho}_B$ were 5.1~kHz and 5.2~kHz, 
respectively, with $\tau_\mathrm{w}=$80~ps and 3.5~mW pump power for clockwise and counterclockwise directions 
of the Sagnac interferometers. 

Next, we performed the entanglement swapping. 
We postselect the detection events of D1 and D2 such that the 
heralded single photons in modes~3 and 4 become temporally indistinguishable. 
We reconstructed the density operator $\hat{\rho}_{\rm{swap}}$ of the photon pairs in modes~1 and 2 by
using the detection events in which the four-fold coincidence among D1, D2, D3 and D4 occurs. 
The four-fold coincidence rate was 28~counts/hour with $\tau_\mathrm{w}=$80~ps 
and the measurement time was 106 hours. 
The reconstructed density operator is shown in Fig.~\ref{fig:matrix}~(c). 
An observed fidelity $F_{\mathrm{swap}}=\bra{\psi^-}\hat{\rho}_{\rm{swap}}\ket{\psi^-}$ and the entanglement of formation~(EOF)~\cite{PhysRevLett.80.2245} 
were $0.84 \pm 0.04$ and $0.82 \pm 0.10$, respectively, 
which indicates that a high-fidelity entanglement swapping is realized  
by using asynchronous polarization entangled photon sources and 
time-resolved coincidence measurement. 
We also estimated the maximized fidelity by the local phase shift as 
$F'_{\mathrm{swap}}=\max_{-\pi\leq\theta\leq\pi}\bra{\psi^-_{\theta}}\hat{\rho}_{\mathrm{swap}}\ket{\psi^-_{\theta}}=0.93\pm0.04$ with $\theta=-0.62$ rad, 
where $\ket{\psi^-_{\theta}}\equiv\ket{\H\V}-e^{i\theta}\ket{\V\H})/\sqrt{2}$. 

\subsection{GHZ state generation}
With a similar setup to the one shown in Fig.~\ref{fig:swappingsetup}, we demonstrated the generation of three-photon GHZ state 
using asynchronous photon sources at telecom wavelengths. 
GHZ state can be generated by the quantum parity check~(QPC)~\cite{PhysRevA.64.062311} on one half of the photon pair forming $\ket{\phi^+}$ and 
a diagonally~(D) polarized ancillary photon~\cite{PhysRevLett.82.1345}. 
For this purpose, we changed the setup in Fig.~\ref{fig:swappingsetup} as follows: 
(1)~We used the photon pair source~B for generating a V-polarized photon pair 
by using only counterclockwise pump beam. A V-polarized photon in mode 3 is transformed to a D-polarized photon by using 
a half wave plate~(HWP). 
(2)~We replaced the HBS by a PBS in order to perform the QPC on photons in modes~3 and 4. 
When we detect a V-polarized photon at D2, a D-polarized photon is heralded in mode 3. 

\begin{figure}[t]
 \begin{center}
 \includegraphics[width=\columnwidth]{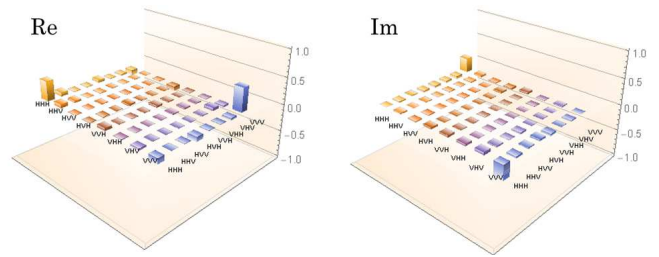}
  \caption{(Color online)~
  The real part and imaginary part of the density operators of $\hat{\rho}_{\rm{GHZ}}$. 
     \label{fig:GHZmatrix}}
 \end{center}
\end{figure}

When the photon in mode~3 is heralded by photon detection 
at D2 and the photon pair is generated in mode~1 and 4, 
the QPC is performed on photons in modes 3 and 4. 
By postselecting the events where all detectors click, 
the polarization state in modes 1, 3' and 4' ideally becomes the 
three-photon GHZ state $\ket{\mathrm{GHZ}}=(\ket{\H\H\H}+\ket{\V\V\V})/\sqrt{2}$ 
with success probability of 1/2. We perform the QST on them 
to reconstruct its density operator $\hat{\rho}_\mathrm{GHZ}$.

In this experiment, we set the pump power to be 5~mW on average 
for achieving higher four-fold coincidence rate. 
The detection rate of the photon pairs generated by the photon source~A and the photon source~B were $1.3\times10^4$ Hz and $7.4\times10^3$ Hz, 
respectively with $\tau_\mathrm{w}=$80~ps. The four-fold coincidence rate was 131~counts/hour and the measurement time was 87~hours. 

The reconstructed density operator~($\hat{\rho}_{\mathrm{GHZ}}$) is shown in Fig.~\ref{fig:GHZmatrix}. 
The imaginary part arises from the non-zero relative phase  between the H-polarized and V-polarized photons. 
We estimated the fidelity maximized by the local phase shift $F_{\mathrm{GHZ}}$ defined by 
$F_{\mathrm{GHZ}}=\max_{-\pi\leq\theta\leq\pi}\bra{\mathrm{GHZ}_{\theta}}\hat{\rho}_{\mathrm{GHZ}}\ket{\mathrm{GHZ}_{\theta}}$, 
where $\ket{\mathrm{GHZ}_{\theta}}\equiv(\ket{\H\H\H}+e^{i\theta}\ket{\V\V\V})/\sqrt{2}$. 
We obtained $F_{\mathrm{GHZ}}=0.70 \pm 0.05$. 
In order to verify that $\hat{\rho}_{\mathrm{GHZ}}$ is a genuine three-photon entangled state, 
we used the witness operator~\cite{PhysRevLett.87.040401, PhysRevLett.92.087902}
\begin{equation}
\mathcal{W}=\frac{\hat{I}}{2}-\ketbra{\mathrm{GHZ}_{\theta}}{\mathrm{GHZ}_{\theta}}, 
\end{equation}
where $\hat{I}$ is the identity operator. 
We obtain $\mathrm{Tr}(\mathcal{W}\hat{\rho}_{\mathrm{GHZ}})=1/2-F_{\mathrm{GHZ}}=-0.20 \pm 0.05 < 0$, 
which shows that $\hat{\rho}_{\mathrm{GHZ}}$ possess genuine three-photon entanglement. 
This result ensures that the multi-photon entangled state can be generated by using fully autonomous sources 
via a time-resolved measurement. 
Such a telecom-band multipartite entanglement source has various applications such as nonlocality test and quantum communication. 

\section{Discussion}
\begin{figure}[t]
 \begin{center}
 \includegraphics[width=\columnwidth]{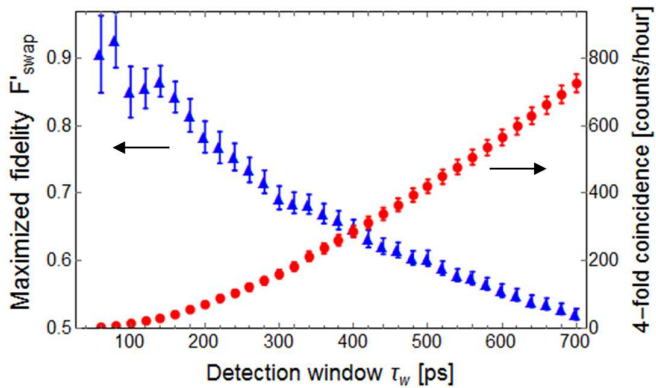}
  \caption{(Color online)~The fidelity of the final state 
  maximized by local phase shift $F'_\mathrm{swap}$~(triangle) and the four-fold 
  coincidence rate~(circle) for various values of $\tau_{\mathrm{w}}$. 
  The four-fold coincidence rate is proportional to $\tau_{\mathrm{w}}^2$ 
for small $\tau_\mathrm{w}$ because 
the two-fold coincidence probability of single photon pair is approximately proportional to 
$\tau_{\mathrm{w}}$ in this region. On the other hand, 
for large $\tau_{\mathrm{w}}$, the four-fold coincidence rate increases linearly 
because the influence of temporally-continuous stray photons becomes dominant in this region. 
\label{fig:width}}
 \end{center}
\end{figure}
In the experiment using asynchronous sources, 
the fidelity of the final state and 
the four-fold coincidence rate depend on 
the width of each detection time window. 
Here we discuss the relation among them. 
In asynchronous source experiment, 
employing small detection time windows is
important  for the following two reasons. 
(1)~For satisfying the temporal indistinguishability of independent SPDC photons~\cite{halder2007entangling}. 
(2)~For suppressing non-negligible temporally-continuous stray photons ~\cite{1608.04909}. 
For investigating the influence of the width of each detection window $\tau_{\mathrm{w}}$, 
we analyzed the experimental data
of entanglement swapping for various values of $\tau_{\mathrm{w}}$ by
post-processing of the common record of timestamps. 
There is a tradeoff between
the four-fold coincidence rate and 
$F'_\mathrm{swap}$. 
In Fig.~\ref{fig:width}, 
we see that the $F'_\mathrm{swap}$ still retains 0.75 $\pm$ 0.02 
even if we set $\tau_{\mathrm{w}}=\tau_{\mathrm{c}}$=230~ps, where 
the four-fold coincidence rate is about 100 counts/hour,  
which is 20 times higher than the previous reported value~\cite{halder2007entangling} 
with the similar fidelity. 
In addition, the four-fold coincidence rate exceeds 500 counts/hour for $\tau_{\mathrm{w}}$=560~ps 
with $F'_\mathrm{swap} > 0.5$, which shows retainment of entanglement~\cite{PhysRevA.54.3824}. 

\section{Conclusion}
In conclusion, we have demonstrated 
the high-fidelity entanglement swapping and the first demonstration of 
generating telecom-band three-photon entangled state by 
using two independent photon pairs generated by SPDC process with cw pump light. 
The observed fidelities are $F_{\mathrm{swap}}=0.84\pm0.04$ and $F_{\mathrm{GHZ}}=0.70 \pm 0.05$ with $\tau_{\mathrm{w}}$=80~ps, 
which are as high as those observed in pulsed~(synchronized) regime. 
In addition we investigated relation among $\tau_\mathrm{w}$, the maximized fidelity and four-fold coincidence rate. 
We revealed that the four-fold coincidence rate becomes 100~counts/hour with the maximized fidelity of 0.75 $\pm$ 0.02 
if we set  $\tau_{\mathrm{w}}=\tau_\mathrm{c}$ for entanglement swapping. 
For further enhancement of the four-fold coincidence rate without degrading the fidelity of the final state, 
the wavelength division multiplexing is effective~\cite{lim2010wavelength, aktas2016entanglement}. 
For instance, when 
the width of emission spectrum of SPDC
is several tens of nanometer, we can utilize several thousand frequency modes with the current filter band width, 
which will drastically improve detection rate of photon pairs 
as long as the linewidth of the pump laser is less than the width of the frequency bin. 
We believe that our results will be useful for many applications for a synchronization-free long-distance quantum communication.

\section*{Aknowledgments}
This work was supported by Core Research for Evolutional Science and
Technology, Japan Science and Technology Agency~(CREST, JST) and 
JSPS Grant-in-Aid for Scientific Research JP16H02214, JP25286077, JP15H03704 and 
JP16K17772. 
RI, TY and NI were supported by JSPS Bilateral Open Partnership Joint Research Projects. YT was supported by 
JSPS Grant-in-Aid for JSPS Research Fellow JP16J05093.

\end{document}